\def\order#1{{\cal O}\left(#1\right)}
\def\sep{ }
\newcommand{\plus}{\phantom{-}}
\def\Ps-{${\rm Ps}^-$}

\documentclass[aps,prl,showpacs]{revtex4}

\usepackage{psfig}

\usepackage{times}
\usepackage{nicefrac}
\usepackage{amsmath}
\usepackage{amsfonts}
\usepackage{amssymb}
\usepackage{amsthm}

\begin{document}
\preprint{Alberta Thy 09-07}

\title{\hfill {\rm Alberta Thy 09-07}\\[4mm]
Positronium-ion decay}

\author{Mariusz Puchalski}

\author{Andrzej Czarnecki}

\affiliation{Department of Physics, University of Alberta,
             Edmonton, AB, Canada T6G 2G7}

\author{Savely G. Karshenboim}
\affiliation{D. I. Mendeleev Institute for Metrology,
St. Petersburg 190005, Russia and Max-Planck-Institut f\"ur Quantenoptik,
Garching 85748, Germany}

\begin{abstract}
 We present a precise theoretical prediction for the decay width of the
bound state of two electrons and a positron (a negative positronium
ion), $\Gamma(\textrm{Ps}^-) = 2.087\,085(12) /\textrm{ns}$. 
We include $O(\alpha^2)$ effects of hard virtual photons as well as
soft corrections to the wave function and the decay amplitude.  An
outcome of a large-scale variational calculation, this is the first
result for second-order corrections to a decay of a three-particle bound
state.  It will be tested experimentally in the new positronium-ion 
facility in Garching in Germany.
\end{abstract}

\maketitle


Positronium ion (\Ps-), consisting of two electrons and a positron, is
the only known three-body bound state free from nucleons.  
Its existence was predicted by Wheeler in 1946
\cite{wheeler1946} and confirmed experimentally by Mills in 1981
\cite{mills1981}.  
Only the ground state is stable  against a dissociation into positronium
and an electron (see \cite{FleischerPhD} for an extensive
review of its properties and references).
Electron-positron annihilation limits the \Ps-
lifetime  to about half a nanosecond, as first reported in
\cite{mills1983}.  Here we determine relativistic and radiative
corrections to the annihilation in a three-body bound state and
predict the \Ps-
decay rate with a 6 parts per million precision,
\begin{equation}
\Gamma(\makebox{\Ps-}) = 2.087\,085(12)\, \textrm{ns}^{-1}.
\label{eq:gamma_value}
\end{equation}

What makes the \Ps- ion particularly interesting is that its theory is
very clean, albeit somewhat technically challenging. 
With a very good accuracy all but electromagnetic interactions in \Ps-
can be neglected.  Also the charge distribution of constituents is
well known (point-like), unlike in atoms and ions containing nuclei.  
Quantum electrodynamics (QED) suffices to describe all
properties of \Ps-.  

On the other hand, \Ps- is a three-body system and thus its wave
function is not known analytically even in the non-relativistic
approximation.  This complicates theoretical investigations but also
provides an opportunity to develop and test advanced computational techniques.
Those new methods are important for  other systems such as the
hydrogen ion, the molecule $H_2^+$, and the helium atom.

\begin{figure}[htb]
\begin{tabular}{cc}
\psfig{figure= 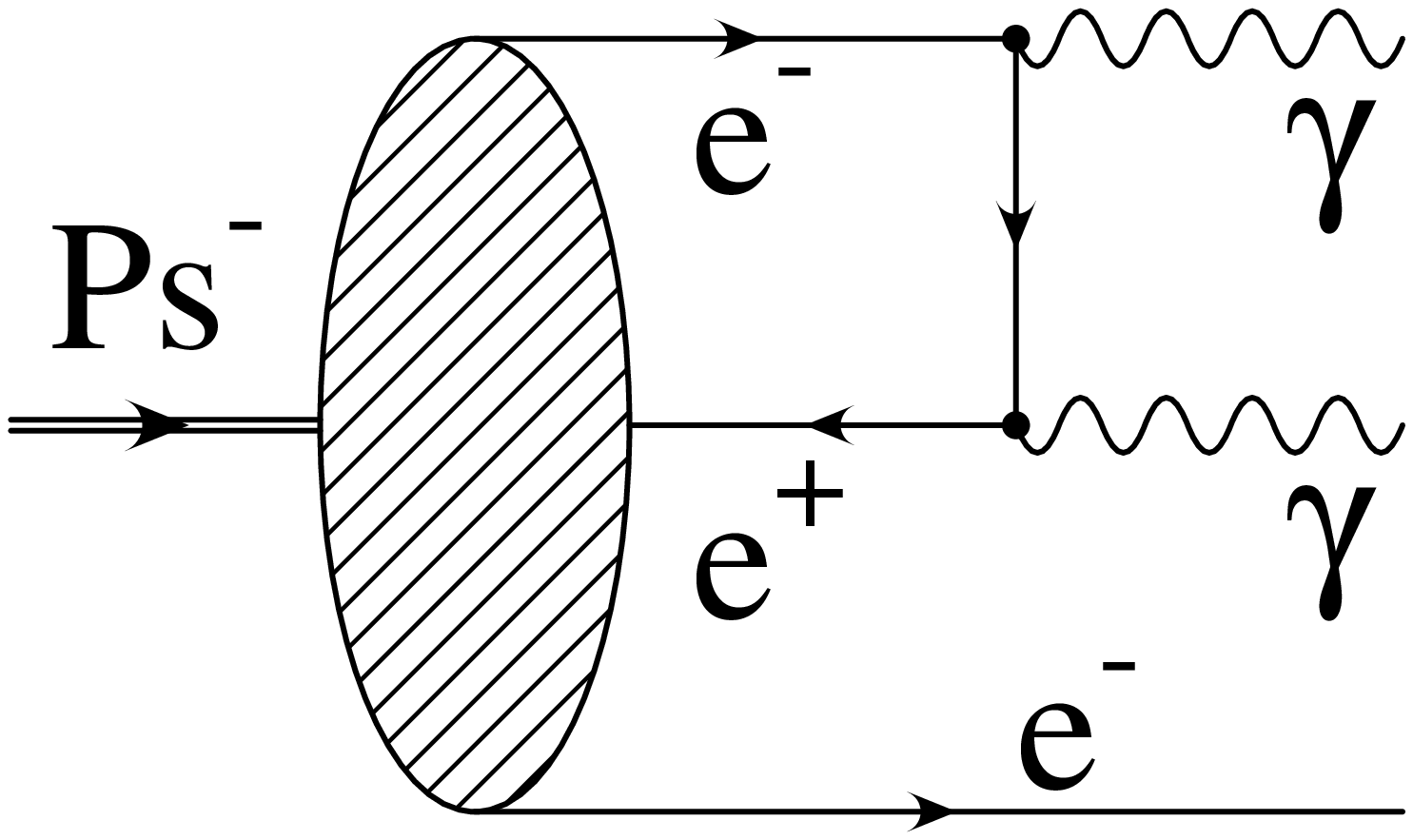,width=40mm} & 
\psfig{figure= 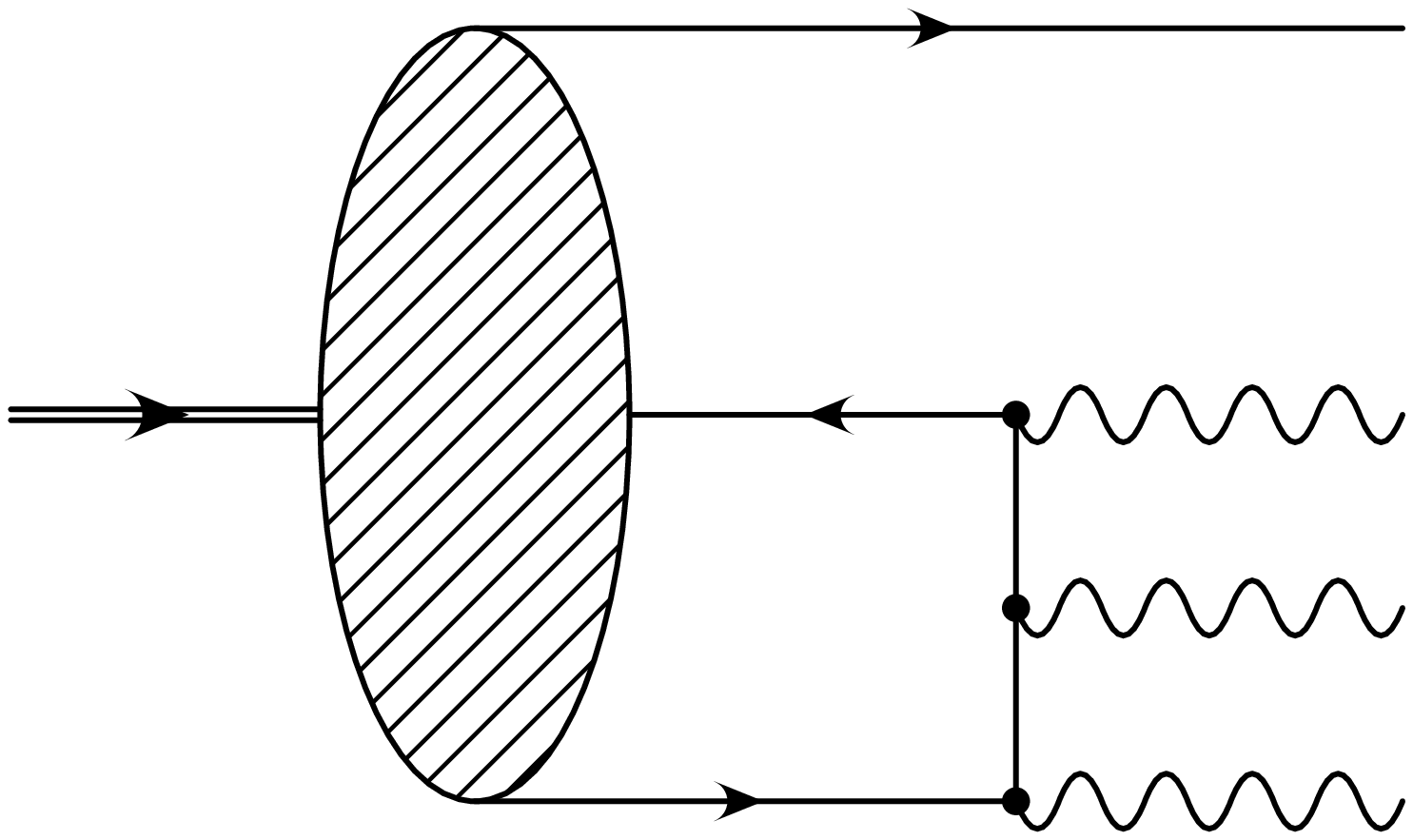,width=40mm}
\\
(a) & (b) 
\end{tabular}
\caption{The main decay channel of \Ps- (a), and an example of 
a correction to it, the three-photon annihilation (b).}
\label{fig:topologies}
\end{figure}

The $e^+e^-$ annihilation proceeds fastest when the pair is in a
spin-singlet state, like para-positronium (pPs), in which case two
photons can be produced (see Fig.~\ref{fig:topologies}(a)).  If the
pair is a spin triplet, like ortho-positronium (oPs), the decay results in
an odd number of photons, Fig.~\ref{fig:topologies}(b).
Interestingly, unlike ortho-positronium, \Ps- can also decay into a
single photon.  However, this channel is very rare 
\cite{Kryuchkov94,Frolov94}: all three constituents
have to overlap to transfer momentum to the non-annihilating electron.
The three-photon decay is much more likely, but still much slower than
the spin-singlet two-photon process.

The spatial wave function of \Ps- is symmetric with
respect to the two electrons.
For the total wave function to
be antisymmetric, the two electrons must be in the spin-singlet state.
It is convenient to think of \Ps- as consisting of a positronium
core and a loosely bound electron \cite{bhatia1983}.  This picture
reveals the main features of the \Ps- lifetime.  
When $e^+$ meets one of the $e^-$, the odds are about one
in four that their spins form a singlet.  Thus \Ps- lives about four
times longer than pPs.

A variational determination of the \Ps- wave function \cite{bhatia1983}
confirms this elegant argument.  Furthermore, if this decay is so
similar to that of pPs, the same $\order{\alpha}$ corrections apply
\cite{Harris57}.  In the same order, also the three-photon
annihilation must be accounted for \cite{Ore1949}.  Together, this led to the
theoretical prediction for the \Ps- decay width
\cite{bhatia1983},
\begin{equation}
\Gamma^{1983}_{\rm th} = 2.086(6) \, {\rm ns}^{-1} ,
\label{eq:oldTh}
\end{equation}
where the size of the $\order{\alpha}$ corrections was used to
estimate the uncertainty \cite{fleischer2006} 
(see also \cite{Ho83,Krivec93,Frolov99}).

Recent measurement \cite{fleischer2006} agrees with this prediction
and approaches its precision,
$\Gamma_{\rm exp} = 2.089(15) \, {\rm ns}^{-1}$.
It is anticipated that the new intense source of positrons at the
Garching reactor FRM-II  will be used to decrease
the experimental error by a factor of 4 or 5, below the
uncertainty in Eq.~(\ref{eq:oldTh}).  

Motivated by this effort, we undertook to improve the
theoretical precision by determining all $\order{\alpha^2}$
effects.   \Ps-,  a non-relativistic bound state, is
well described by the Schr\"odinger equation. Its leading-order decay rate
is
\begin{equation}
\Gamma_{0} = 2 \pi m_e  \alpha^5
 \left\langle \delta^3(r_{12})\right\rangle ,
\label{eq:gamma_lo}
\end{equation}
where $m_e$ is the electron mass, $r_{12}$ is the distance between
the positron and the electron which annihilates, and the mean value
refers to the ground state \Ps- wave function
\begin{equation}
\Psi = \psi(r_{12},r_{13},r_{23})\, \chi(1,2,3), \quad \chi =
\uparrow_1 
{\downarrow_2 \uparrow_3 - \uparrow_2 \downarrow_3
\over
\sqrt{2} }
.
\label{eq:wfunc_nr}
\end{equation}
Throughout this paper we use $1/\alpha m_e$, $\alpha m_e$, and $\alpha^2m_e$
as units of length, momentum, and energy (we also set
$c=\hbar = 1$, except in the last Eq.~(\ref{eq:finEq})).  Thus, 
$\left\langle \delta^3(r_{12})\right\rangle$ in
Eq.~(\ref{eq:gamma_lo}), as well as all mean values to follow, are  
dimensionless.  

Relativistic effects, spin of the electron, and short-distance
exchanges of photons with a virtuality $\order{m_e}$ are not accounted
for by the Schr\"odinger equation, which includes only Coulomb
potentials among the three constituents.  Like in other
non-relativistic systems \cite{Caswell:1986ui}, these additional
effects can be treated as perturbations and organized in a series in
$\alpha$,
\begin{eqnarray}
\Gamma &=&  \Gamma_{0} \, 
\left[ 1 
+ \alpha \, A 
+\alpha^2 \left( 2 \ln {1\over \alpha}  + B\right)
 - \frac{3\alpha^3}{2 \pi}  \ln^2 {1 \over \alpha}
 + \ldots \right].
\label{eq:gamma_expansion}
\end{eqnarray}
The first-order correction $A$, already discussed, includes
corrections to the two-  and the
three-photon channels, 
\begin{equation}
A\equiv  A^{2 \gamma} + A^{3 \gamma}, 
\qquad
A^{2 \gamma} =   \frac{\pi}{4} - \frac{5}{\pi}, 
\qquad
A^{3 \gamma} = 
\frac{4\pi}{3}
- {12\over \pi}.
 \label{eq:gamma_a}
\end{equation}
Some authors (e.g.~\cite{bhatia1983}) hint at additional
$\order{\alpha}$ effects but in our opinion none other exist
at this order.

In the next order, four photons contribute
\cite{Adachi:1990nj}, corrections $\order{\alpha}$ 
must be included in the three-photon decay
\cite{Caswell:1976nx,Adkins96}, and $\order{\alpha^2}$ in the
two-photon decay \cite{Czarnecki:1999gv,Czarnecki:1999ci,Adkins:2001},
\begin{equation}
B = B^{4 \gamma} + B^{3\gamma} + B^{2 \gamma}. 
\label{eq:gamma_a2}
\end{equation}
The last term is the focus of this paper.  It is a sum of several
effects: square $B_{\rm squared}$ of the $\order{\alpha}$ correction
$A^{2\gamma}$; hard-photon corrections $B_{\rm hard}$ to the
$e^+e^-\to \gamma\gamma$ process; and soft corrections to the
annihilation amplitude $B_{\rm aa}$ and the wave function $B_{\rm
wf}$:
\begin{eqnarray}
B^{2 \gamma} &=& B_{\rm squared}  + B_{\rm hard} 
+ B_{\rm aa}  + 
B_{\rm wf} ,
\nonumber \\
B_{\rm squared}  &=& \left( {5\over 2\pi} -
\frac{\pi}{8} \right)^2, 
\label{eq:gamma_squared}
\\
B_{\rm hard}  &\equiv &B_{\rm hard}^{\rm fin}-{1\over 2\epsilon},\quad 
B_{\rm hard}^{\rm fin}  =  - \frac{40.46(30)}{\pi^2} ,
\label{eq:gamma_hard}
\\
B_{\rm aa}  &=& \frac{1}{3}. 
\label{eq:gamma_ampl}
\end{eqnarray}
All corrections which affect only the annihilation amplitude have
already been computed in the context of the pPs decay.  
Since they do
not depend on the particular bound state, they apply to the
present analysis without changes.

The correction to the wave function $B_{\rm wf}$, sensitive to the
three-body dynamics, is the most challenging.  As we will see below,
it is divergent and cancels the divergence in $B_{\rm hard}$.  The
term $-2\alpha^2 \ln\alpha$ in Eq.~(\ref{eq:gamma_expansion}) is a
remnant of those divergences.  In order to regularize divergences, we   
work in
$d=3-2\epsilon$ spatial dimensions.  Thus, the non-relativistic
Coulomb Hamiltonian becomes
\begin{eqnarray}
H_0 &=& \sum_{a}\,\frac{\vec p_a^{\,2}}{2} + V, 
\\
V &\equiv & - \biggl[\frac{1}{r_{12}}\biggr]_{\epsilon} -
\biggl[\frac{1}{r_{13}}\biggr]_{\epsilon} +
\biggl[\frac{1}{r_{23}}\biggr]_{\epsilon} 
\equiv
\sum_{a<b}\,z_{ab}\,\biggl[\frac{1}{r_{ab}}\biggr]_{\epsilon},
 \nonumber
\\
\biggl[\frac{1}{r}\biggr]_{\epsilon} &\equiv & 
\frac{\pi^{\epsilon - {1\over 2}}
\Gamma \left({1\over 2} - \epsilon\right)}
{r^{1-2 \, \epsilon}}, 
\label{eq:hamiltonian_a2}
\end{eqnarray}
where  $\vec p_{a}$, $\vec r_{ab} \equiv \vec r_{a} - \vec r_{b}$ are
momenta and relative distances of the positron 1 and electrons
2, 3.

The wave function correction arises due to relativistic effects, which
are treated as a perturbation and described by the Breit Hamiltonian,
\begin{eqnarray}
H^{(4)} &=& \alpha^2 \left( H^{(4)}_1 + \sum_{a<b} H^{(4)}_{2\,ab}
 \right)
\label{hamiltonian_a4} 
\\
H^{(4)}_1 &=& -\sum_a \frac{p^{\,4}_a}{8}   
- \sum_{a<b}z_{ab}\left\{ \, \pi\,  \delta^d(r_{ab})
+\frac{1}{2} p_a^i\,\left[ {1\over r_{ab}} \right]_{\epsilon} 
\left(\delta^{ij}+(d-2)\frac{r^i_{ab}\,r^j_{ab}}{r^2_{ab}}
\right)\, p_b^j 
 \right\} 
\nonumber \\
H^{(4)}_{2\,ab} &=& \frac{\pi z_{ab}}{4d}\,[ \sigma^i_a, \sigma^j_a ][
\sigma^i_b, \sigma^j_b ]
\,\delta^d(r_{ab}) 
\label{h2ab}
\end{eqnarray}
where the Pauli matrices are labeled with the number of the
fermion  they are acting on.  It is convenient to
evaluate separately the spin-independent part
$H^{(4)}_1$, and the spin-dependent part $H^{(4)}_2$, $B_{\rm
wf}  \equiv B_{H1}  + B_{H2} +{1\over 2\epsilon} $.

The effect of this perturbation is the following replacement in the
formula for the decay rate, Eq.~(\ref{eq:gamma_lo}),
\begin{equation}
\left\langle \delta^3 (r_{12}) \right \rangle
 \to 2\left\langle \delta^d (r_{12})
{1\over (E-H_0)'} H^{(4)} \right\rangle 
\equiv
\alpha^2 
\left( B_{H1} 
+  B_{H2} + {1\over 2\epsilon}\right)
\left\langle \delta^3 (r_{12}) \right\rangle.
\label{eq:defBH}
\end{equation}
Here $\frac{1}{(E - H)^{'}}$ is the Green's function of the
lowest-order Schr\"odinger equation and the prime indicates the
exclusion of the ground state.
The appearance of divergences is the main obstacle in the evaluation
of this correction.  They originate from $r_{12}\to 0$ (ultraviolet
limit), where the Breit Hamiltonian is not a valid description of the
dynamics.  Indeed, when one accounts for the hard photons,
Eq.~(\ref{eq:gamma_hard}), divergences cancel.

In analogy with the earlier work on positronium and helium
\cite{Czarnecki:1999mw,Pachucki98,yelkhovsky2001,pachucki:022512}, 
we rewrite the matrix element
in Eq.~(\ref{eq:defBH}) such that the divergences appear only in the
coefficient of one operator, namely $\delta^3(r_{12})$.  To this end,
we rewrite \cite{pachucki:022512} the delta-function as
\begin{equation}
4 \, \pi \, \delta^d (r_{12}) = 4 \, \pi \, \widetilde{\delta}^d
(r_{12}) 
+ \biggl\{H_0 - E,\biggl[ \frac{1}{r_{12}}\biggr]_{\epsilon}\biggr\}.
\label{eq:regularized_delta}
\end{equation}
This equation implicitly defines $\widetilde{\delta}^d$,  less
singular than $\delta^d$.  The most singular part is in the
anticommutator in the  second term.  This term cancels the
Green's function, 
$(E-H_0) \frac{1}{(E - H_0)^{'}} 
=I - \left| \Psi \right\rangle \left\langle \Psi \right|$,
where $I$ is the identity operator.
Hence,  divergences
appear only in first-order elements and are easier to 
extract.  

In the spin-independent part we find
\begin{eqnarray}
2\pi 
\left\langle \delta^3(r_{12}) \right\rangle 
\left( B_{H1}  + {1\over 4\epsilon} \right)
= 
\left\langle 4 \, \pi  \delta^d(r_{12})
\,
\frac{1}{(E - H_0)^{'}}\, H_1^{(4)} \right\rangle
=
{1\over 4} 
\sum_{i=1}^{21} v_i
+ {1\over E^2}v_{1} v_{22} +v_{23}
+ {\pi \over 2\epsilon}\left\langle \delta^3(r_{12})\right\rangle ,
\end{eqnarray}
from which we can determine the value of $ B_{H1}$ in terms
of the ground-state mean values listed in Table
\ref{tab:tableV}.  
\begin{table}[htb]
\caption{\label{tab:tableV}
Operators affecting the \Ps- wave function and their  ground state mean
values.  We denote $V_{12}\equiv V + 1/r_{12}$.}
\begin{ruledtabular}
\begin{tabular}{llr}
$i$ & Operator $O_i$  & $v_i=\langle O_i \rangle $ \\
\hline \\
1 & ${E^2 / r_{12}}$   & $0.023~327~6
$ \\
2 & ${V_{12}^2 / r_{12}}$ &$ 0.033~945~0
$ \\
3 & ${2Ep_3^2 / r_{12}}$ & $-0.014~986~7
$ \\
4  & $-{2EV_{12} / r_{12}}   $  &$ -0.015~844~8
$ \\
5  &  $ -2{p_3^2 V/ r_{12}} $  & $0.090~907~3
$ \\
6  &  $ -{p_3^4 / r_{12}} $  & $-0.029~850~5
$ \\
7  &  $ -{4\pi\delta^3(r_{13})/ r_{12}}$  & $-0.051~014~0
$ \\
8  &  $ {4\pi\delta^3(r_{23})/ r_{12}} $  & $0.001~794~7
$ \\
9  &  $ {3E/ r_{12}^2} $  & $-0.219~554~9
$ \\
10  &  $-{3V_{12} / r_{12}^2} $  &$ 0.022~092~3
$ \\
11  &  $-(\vec p_1 \times  \vec p_2 )^i
           (1 / r_{12})(\vec p_1 \times \vec p_2)^i  $  &
$-0.001~809~5
$ \\
12  &  $-({4\, \pi}/{3})\, \delta^3(r_{12})\,p_3^2 $  &
$-0.003~350~2
$ \\
13  &  $ -\, p_3^i\left( 3\,{r_{12}^i\, r_{12}^j}/{r_{12}^5}
-
{\delta^{ij}}/{r_{12}^3}\right)p_3^j/2$  & $-0.000~275~1
$ \\
14  &  $  {r_{12}\cdot r_{13} /( 2\, r_{12}^3 r_{13}^3 )} $  &
$-0.000~562~9
$ \\
15  &  ${r_{12}\cdot r_{23} / (2\, r_{12}^3 r_{23}^3) } $  &
$-0.001~413~3
$ \\
16  &  $ -p_1^2 (1/ r_{12}) p_3^2 $  & $-0.037~118~4
$ \\
17  &  $ -p_2^2 (1/ r_{12}) p_3^2 $  & $-0.008~951~8
$ \\
18  &  $  - \sum_a \, p_a^i (1 / r_{12}^2)p_a^i/2$  &
$-0.077~005~7
$ \\
19  &  $ 2 \,\sum_{a<b} \,z_{ab}\,
p_a^i$ & \\
& $\left({\delta^{ij}}/{r_{ab}}+{r^i_{ab}\,r^j_{ab}}/{r^3_{ab}}
\right)/r_{12}\, p_b^j   $  & $0.296~062~9
$ \\
20  &  $2\pi \delta^{3}(r_{12}) $  & $0.130~270~5
$ \\
21  &  $ 2\, P \left({1}/{r^3}\right) $  & 
$0.014~113~8
$ \\
22  &  $ H_1^{(4)} $  &$ -0.072~738~1
$ \\
23  &  $ 4 \,\pi\, \widetilde{ \delta^3}(r_{12}) \,
\frac{1}{(E - H_0)^{'}}\, H_1^{(4)} $  &$ 0.178~732~5
$ \\
24  &  $ \sum_{a < b}\,   4\pi\, A_{ab} \widetilde{\delta}^3(r_{12})
\,\frac{1}{(E - H_0)^{'}}\, \pi\, \widetilde{\delta^3}(r_{ab})  $  &
$0.334~788~9
$ \\
25  &  $ 2\pi\delta^3(r_{23}) $  & $0.001~074~4$
\end{tabular}
\end{ruledtabular}
\end{table}
Among them, the regularized cubic operator is defined by Eq. (1.5) in
Ref.~\cite{pachucki:022512}. 

In the spin-dependent part, the effect of Pauli matrices in
$H^{(4)}_{2ab}$, Eq.~(\ref{h2ab}), is evaluated with the spin
wave function in Eq.~(\ref{eq:wfunc_nr}) and represented by constants
$A_{ab}$ for each pair of fermion lines:  
$A_{12} = -2 - 6 \epsilon$, $A_{13} = -A_{23} = -2$.  We keep
$\epsilon$ only in the coefficient of the divergent part. 
After this simplification of spins, we find
\begin{eqnarray}
2\pi 
\left\langle \delta^3(r_{12}) \right\rangle 
\left( B_{H2}  + {1\over 4\epsilon} \right)
= 
\sum_{a < b} \, \left\langle 4 \,\pi\,
\delta^d(r_{12}) \frac{1}{(E - H_0)^{'}}\,
\pi\, A_{ab}\, \delta^d(r_{ab})  \right\rangle. 
\label{eq:gamma_H2}
\end{eqnarray}
In terms of the operators in Table \ref{tab:tableV},  using the
symmetry $\vec r_2 \leftrightarrow \vec r_3$ and the virial
identity $2E = \left\langle V \right\rangle$, we get
\begin{eqnarray}
2\pi 
\left\langle \delta^3(r_{12}) \right\rangle 
 B_{H2}
=
 {1\over 2} \left( -v_7 - v_8 -v_{14} - v_{15} +v_{18} \right)
+ {1\over 6} \left( v_9 + v_{10} \right) 
+{8E+5\over 4} v_{20}
+ {1\over 4} v_{21} 
+v_{24} 
+ {v_{1}\over E^2}(v_{25} -2v_{20} ).
\end{eqnarray}

The numerical values in Table \ref{tab:tableV} are obtained with a
variational method.  The trial wave function is expanded in
a 1000-element set of exponential functions \cite{korobov2000}
\begin{equation}
 \phi(r_{12},r_{13},r_{23}) = \sum_{k=1}^2\,\sum_{i=1}^{N_k} \,
 d_{ki}\,e^{-a_{ki} r_{12} - b_{ki} r_{13} - c_{ki} r_{23}}
 + (r_{12} \leftrightarrow r_{13}),
\end{equation}
where $a$, $b$, $c$ are chosen randomly, with a 
homogeneous distribution,  from two $k$ sets
defined by variational boundary conditions $A_{1k} \leq a_{ki}
\leq A_{2k}$, $B_{1k}\leq b_{ki} \leq B_{2k}$, $C_{1k}\leq c_{ki}
\leq C_{2k}$. 
Two (or more) sets allow one to match the behavior of
the wave function at various
distance scales and  improve  accuracy.
We
found the non-relativistic energy value, 
$E = -0.262\,005\,070\,232\,980(1)$,
that agrees with an even more accurate earlier result \cite{Frolov2006}. 
Previously obtained mean values of 
$\delta^3(r_{ab})$ \cite{Frolov99,drake2005}, 
and non-singular products of
$1/r_{ab}$  \cite{Frolov99}
are also confirmed.  Finally, the mean value of the
spin-independent part of the Breit
Hamiltonian $H^{(4)}_1$ agrees with Ref.~\cite{drake2005}.   
Crucial for the decay is the mean value of the
delta-function, obtained using the representation of Ref.~\cite{Drachman81}, 
\begin{equation}
\left\langle \delta^3(r_{12}) \right\rangle = 0.020\,733\,198\,005\,1(2).
\label{eq:delta12}
\end{equation}
This value agrees with the one found in \cite{drake2005} and somewhat improves
its accuracy.     

\begin{table}[htb]
\caption{\label{tab:table1} Corrections to the width of \Ps-.}
\begin{ruledtabular}
\begin{tabular}{ll}
  Correction  &  Value \\
\hline
  $\alpha  \sep A^{3 \gamma}$               &  $\plus 0.002~693~245  $\\
  $\alpha  \sep A^{2 \gamma}$               &  $ -0.005~882~770  $\\
  $-2\alpha^2 \sep  \ln \alpha $          &  $\plus 0.000~524~019  $\\
  $\alpha^2  \sep B^{4 \gamma}$               &  $\plus 0.000~001~480  $\\
  $\alpha^2 \sep B^{3 \gamma}$               &  $ -0.000~064~352  $\\
  $\alpha^2 \sep B_{\rm squared} $          &  $\plus 0.000~008~652  $\\
  $\alpha^2 \sep B_{\rm hard}^{\rm fin} $        &  $ -0.000~218~3(34)$\\
  $\alpha^2 \sep B_{\rm aa} $                  &  $\plus 0.000~017~750  $\\
  $\alpha^2 \sep B_{\rm H1} $          &  $\plus 0.000~078~366  $\\
  $\alpha^2 \sep B_{\rm H2} $          &  $\plus 0.000~122~185  $\\
  $3\alpha^3 \sep  \ln^2 \alpha/(2 \pi)$&  $ -0.000~004~491  $\\
  $ \sep 2.5(2.5)\alpha^3 \ln \alpha$      &  $ -0.000~004~8(48)  $\\
\hline
 Total $C$ & $-0.002~729~0(59)$\\
\end{tabular}
\end{ruledtabular}
\end{table}
For the new evaluation of the \Ps- decay rate we use
$\alpha=1/137.03599911(46)$ and the atomic unit of time $\alpha^2
m_e c^2/\hbar = 10^{17}\, {\rm s}/2.418 884 326 505(16)$ \cite{Mohr05}.
Our final result in Eq.~(\ref{eq:gamma_value}) is obtained using
\begin{equation}
\Gamma( \makebox{\Ps-} ) = 2\pi {\alpha^5 m_e c^2 \over \hbar} (1+C) 
\left\langle \delta^3(r_{12}) \right\rangle,
\label{eq:finEq}
\end{equation}
where the correction $C$ is given in Table \ref{tab:table1}, and we
use Eq.~(\ref{eq:delta12}).  The last two corrections listed in Table
\ref{tab:table1} refer to the third order in $\alpha$.  The leading
quadratic logarithm was found in \cite{DL} and is valid for
positronium atoms as well as for the ion.  The linear log (the last
correction) has not yet been calculated for
\Ps-.  However, it is known for pPs and oPs
\cite{Kniehl:2000dh,Hill:2000qi,Melnikov:2000fi}.  We expect its value
for \Ps- to be close to that for pPs and use the latter as an
estimate.  We assign this correction a 100\% uncertainty, which also
conservatively estimates non-logarithmic higher-order effects
\cite{Penin:2003jz}.

Note that Ref.~\cite{drake2005} includes a prediction of
$\Gamma(\makebox{\Ps-} \to \gamma\gamma)$ with a seemingly higher
precision than ours.  That result, however, does not include any
corrections beyond the tree level (this corresponds to setting $C=0$
in our Eq.~(\ref{eq:finEq})) and its error estimate includes only the
numerical uncertainty of the variational calculation in
\cite{drake2005}.

Another experimentally interesting quantity is  the branching ratio of
the three-photon decay.  We find
\begin{eqnarray}
{\rm BR}(\makebox{\Ps-}\to \gamma\gamma\gamma)
&\equiv &
{\Gamma(\makebox{\Ps-}\to \gamma\gamma\gamma)\over \Gamma(\makebox{\Ps-})}
\nonumber \\
&= &\alpha \left[  A^{3\gamma}
 + \alpha\left( B^{3\gamma}-A A^{3\gamma}\right)
-{7\over 3}A^{3\gamma} \alpha^2 \ln {1\over \alpha} + \ldots \right]
= 0.002~635~8(8).
\end{eqnarray}
The uncertainty is due to the unknown $\order{\alpha^2}$ corrections
to the decay \Ps-$\to \gamma\gamma\gamma$.  Only the logarithmic term
is known in this order \cite{Caswell79}, and we take half of its value
to estimate the uncertainty.

The structure of corrections found in this study confirms the picture
of \Ps- as an electron loosely interacting with a positronium core
\cite{bhatia1983}.  The mean value  in
Eq.~(\ref{eq:delta12}) is very close to that obtained with
a neutral positronium, neglecting the second electron,
$1/(16\pi)=0.01989$.  Also, in the sum of all effects in Table 
\ref{tab:table1}, there is a significant cancellation between the hard
effects $B_{\rm hard}$ and the soft ones $B_{\rm H1}+B_{\rm H2}+B_{\rm
aa}$, also observed in positronium \cite{Czarnecki:1999gv}.  It would
be interesting to understand the origin of this
cancellation, which for now remains an open question.

The accuracy we have obtained for the decay rate is 6 parts per
million, about 500 times better than the previous best prediction,
Eq.~(\ref{eq:oldTh}). 
Further progress in the theory of the \Ps- decay requires 
the logarithmic term $\order{\alpha^8 \ln \alpha}$ and
improved hard corrections $\order{\alpha^7}$.
However, the accuracy obtained in the present paper is sufficient for
the foreseeable future.  It exceeds the anticipated accuracy of
Garching measurements by about a factor of 200.  We are thus prepared
for the new data and are looking forward to this intriguing test of
three-body bound-state QED.

We thank K. Pachucki and D. Habs for helpful discussions.  
This work was supported by the Alberta Energy Research Institute,
under the COURSE Program, by Science and Engineering Research Canada,
RFBR (grants \# 06-02-16156 and \# 06-02-04018) and DFG (grant GZ 436
RUS 113/769/0-2).


\end{document}